\documentclass[sigconf]{acmart}
\AtBeginDocument{%
  }

\usepackage{graphicx}
\usepackage{subcaption}
\usepackage{tabularx}
\usepackage{multirow}
\usepackage{booktabs}
\usepackage{colortbl}
\usepackage[table]{xcolor}

\definecolor{Gray}{gray}{0.85}
\newcolumntype{o}{>{\centering\arraybackslash\columncolor{Gray}}X}
\newcolumntype{R}{>{\raggedleft\arraybackslash}X}

\copyrightyear{2026}
\acmYear{2026}
\setcopyright{cc}
\setcctype{by}
\acmConference[L@S '26] {Proceedings of the Thirteenth ACM Conference on Learning @ Scale}{June 29--July 3, 2026}{Seoul, Republic of Korea.}
\acmBooktitle{Proceedings of the Thirteenth ACM Conference on Learning @ Scale (L@S '26), June 29--July 3, 2026, Seoul, Republic of Korea}
\acmISBN{979-8-4007-2293-6/2026/06}
\acmDOI{10.1145//3774398.3811613}

\begin{document}

%
% \title{Scalable LLM-based Coding of Dialogue in Healthcare Simulation: Balancing Accuracy, Speed and Environmental Impact}
\title[Scalable LLM-based Coding of Team Dialogue in Healthcare Simulation]{Scalable LLM-based Coding of Dialogue in Healthcare Simulation: Balancing Coding Performance, Processing Time, and Environmental Impact}
%\title{Scalable LLM-Based Dialogue Coding for Healthcare Simulation: Accuracy, Speed, and Environmental Trade-offs}
% TODO, Use the same words in RQs and title.

%%
%% The "author" command and its associated commands are used to define
%% the authors and their affiliations.
%% Of note is the shared affiliation of the first two authors, and the
%% "authornote" and "authornotemark" commands
%% used to denote shared contribution to the research.
% \author{Ben Trovato}
% \authornote{Both authors contributed equally to this research.}
% \email{trovato@corporation.com}
% \orcid{1234-5678-9012}
% \author{G.K.M. Tobin}
% \authornotemark[1]
% \email{webmaster@marysville-ohio.com}
% \affiliation{%
%   \institution{Institute for Clarity in Documentation}
%   \city{Dublin}
%   \state{Ohio}
%   \country{USA}
% }

\author{Kiyoshige Garces}
\affiliation{%
  \institution{RMIT University}
  \institution{Monash University}
  \city{Melbourne}
  \country{Australia}}
\email{oscar.garcesaparicio@monash.edu}

\author{Gloria Milena Fernandez-Nieto}
\affiliation{%
  \institution{Monash University}
  \city{Melbourne}
  \country{Australia}}
\email{gloriamilena.fernandeznieto@monash.edu}

\author{Linxuan Zhao}
\affiliation{%
  \institution{Adelaide University}
  \city{Adelaide}
  \country{Australia}}
\email{linxuan.zhao@adelaide.edu.au}

\author{Sachini Samaraweera}
\affiliation{%
  \institution{Monash University}
  \city{Melbourne}
  \country{Australia}}
\email{sachini.samaraweera@monash.edu   }

\author{Dragan Gašević}
\affiliation{%
  \institution{Monash University}
  \city{Melbourne}
  \country{Australia\\}
  \institution{The University of Hong Kong}
  \city{Hong Kong SAR}
  \country{China}
  }
\email{dragan.gasevic@monash.edu}

\author{Roberto Martinez-Maldonado}
\affiliation{%
  \institution{Monash University}
  \city{Melbourne}
  \country{Australia}}
\email{roberto.martinezmaldonado@monash.edu}

\author{Vanessa Echeverria}
\affiliation{%
  \institution{RMIT University}
  \city{Melbourne}
  \country{Australia\\}
\institution{ESPOL University}\city{Guayaquil}
  \country{Ecuador}}
\email{vanessa.echeverria@rmit.edu.au}

%%%%%%%%%%%%%%%%% Authors  %%%%%%%%%%%%%%%%%

%%
%% By default, the full list of authors will be used in the page
%% headers. Often, this list is too long, and will overlap
%% other information printed in the page headers. This command allows
%% the author to define a more concise list
%% of authors' names for this purpose.
\renewcommand{\shortauthors}{K. Garces, G. Fernandez-Nieto, L. Zhao, S. Samaraweera, D. Gasevic, R. Martinez-Maldonado, V. Echeverria.}

%%
%% The abstract is a short summary of the work to be presented in the
%% article.
\begin{abstract}
Extensive research shows that dialogue, the interactive process through which participants articulate and negotiate their thinking, plays a central role in constructing shared understanding, coordinating action, and shaping learning outcomes in teams. Analysing dialogue content has been central to advancing team learning theory and informing the design of computer-supported collaborative learning (CSCL) environments, yet this progress has depended on labour-intensive qualitative coding. Large language models (LLMs) offer new possibilities for automating dialogue analysis and for strengthening the dialogue layer within emerging multimodal learning analytics approaches, with recent studies demonstrating that they can approximate human coding through zero- and few-shot prompting. However, most prior work has focused on replicating human coding accuracy for research purposes, rather than addressing a more educationally consequential question: how can we design prompts that allow an LLM to label team dialogue accurately and fast enough to be useful in real settings, such as in-person healthcare simulations, where results must be returned quickly and computational cost and sustainability also matter? 
Thus, this paper investigates how prompt design and batching strategies can be optimised to balance coding accuracy, processing time, and environmental impact in team-based healthcare simulation debriefing. Using a dataset of $11,647$ utterances coded across six team learning dialogue constructs (codes), we compared four prompt designs across varying batch sizes, evaluating coding performance, processing time, and energy consumption, as well as the trade-offs between these metrics. Results indicate that increasing batch size improves speed and reduces energy use, but negatively impacts coding performance. Beyond demonstrating the feasibility of LLM-based qualitative analysis, this study offers practical guidance for scaling dialogue analytics in contexts where timeliness, privacy, and sustainability are critical.

%Qualitative coding is time-consuming and labour-intensive, yet it remains a key method for analysing and evaluating learner behaviours in educational settings. Large language models (LLMs) have emerged as a promising tool to automate this process, and empirical work has begun to evaluate their accuracy and capability in replicating human coding through techniques such as few-shot and zero-shot prompting — often using cloud-based services. However, little is known about how to apply and scale these models when practical constraints come into play, such as the need for local deployment (e.g., due to data privacy), near real-time processing (e.g., to couple outputs with the timing of a learning activity), and environmental concerns (e.g., the energy consumed by large-scale inference).

% main concepts to develop: ethical, secure, sustainable LLM-based team dialogue coding 
% taking decision for next study - which model, before deployment
% in-the-wild method - real-time or nearly realtime

% papers para el Background - 5 refs

% \keywords{First keyword  \and Second keyword \and Another keyword.}

% Balancing accuracy, speed, and environmental impact.

\end{abstract}
%
%
%

%%
%% The code below is generated by the tool at http://dl.acm.org/ccs.cfm.
%% Please copy and paste the code instead of the example below.
%%

% \ccsdesc[500]{Social and professional topics~Adult education}

%%
%% Keywords. The author(s) should pick words that accurately describe
%% the work being presented. Separate the keywords with commas.
% %% A "teaser" image appears between the author and affiliation
% %% information and the body of the document, and typically spans the
% %% page.
% \begin{teaserfigure}
%   \includegraphics[width=\textwidth]{sampleteaser}
%   \caption{Seattle Mariners at Spring Training, 2010.}
%   \Description{Enjoying the baseball game from the third-base
%   seats. Ichiro Suzuki preparing to bat.}
%   \label{fig:teaser}
% \end{teaserfigure}

\received{16 February 2026}
% \received[revised]{12 March 2009}
% \received[accepted]{5 June 2009}

%%
%% This command processes the author and affiliation and title
%% information and builds the first part of the formatted document.
\maketitle

\section{Introduction}

Dialogue, the interactive process through which participants articulate and negotiate their thinking, sits at the heart of effective teamwork and small-group learning \cite{Fransen01012013}. Across team science and Computer-Supported Collaborative Learning (CSCL), decades of research show that team dialogue shapes shared understanding, coordination, and learning outcomes \cite{Salas2018ScienceTeamwork,HmeloSilverJeong2021CSCLMethods}. Dialogue analysis has been critical for identifying key learning constructs such as explanation, questioning, and co-regulation \cite{Stahl2006GroupCognition}, advancing theory and the design of CSCL environments. Yet, this progress has depended on qualitative coding of dialogue: a rigorous but time-consuming and labour-intensive method \cite{JeongHmeloSilverYu2014CSCLMethods} that limits scale and slows feedback provision to learners or educators \cite{Zhao2024endusers}. As collaborative learning is embedded in formal education \cite{Erdogdu2025CSCLTrends,Yang2022collaboration} and remains essential for developing the collaboration skills required in the workforce \cite{Marshall2024,Papathoma2016}, the CSCL field faces a crucial challenge in efficiently analysing team dialogue at scale and meaningfully supporting learning.

Large language models (LLMs) offer new possibilities for automating dialogue analysis \cite{Wang2025LORA}. Efforts in collaborative learning are not new; earlier work leveraged machine learning and natural language processing (NLP) techniques to classify and model dialogue \cite{Wang2025BERTvsLLM,crowston2012using,devlin2019BERT}. Recent research shows that LLMs can approximate human qualitative coding using zero- and few-shot prompting, substantially lowering technical barriers to implementation \cite{Liu2025Qualitative}. However, most prior studies have focused mainly on replicating human coding accuracy under controlled conditions, often overlooking the practical constraints of real-world deployment \cite{Liu2025Qualitative,wang2024artificial,ZambranoEtAl2023FromnCoderToChatGPT}.

In educational settings that are not fully mediated by online platforms -- such as physical classrooms \cite{Wang2025BERTvsLLM}, specialised learning spaces including laboratories \cite{Martinez2020where,MartinezMaldonado2017DesignEcology}, and high-fidelity simulation rooms \cite{Martinez2017}-- where communication is central to task performance and learning, automated dialogue analysis must do more than approximate human coders. It must operate in authentic, time-sensitive environments where latency, privacy, computational efficiency, and resource use are critical constraints \cite{SamaraweeraEtAl2026LLMHealthcareSimulation,Stenseth2025,Algarni2025}. Robust and efficient dialogue analysis can also strengthen research on communication processes in emerging areas such as multimodal learning analytics, where verbal interaction is integrated with other data streams such as gesture, gaze, or physiological signals \cite{Ouhaichi2023rethinking,Echeverria2025}. Dialogue-driven analytics can provide timely feedback to support reflective practice \cite{Echeverria2024}. Realising these benefits at scale, however, depends on privacy-preserving deployment, and accounts for the computational and environmental impacts associated with large-scale LLM inference \cite{GasevicDawsonSiemens2015LearningAnalytics,ChoiEtAl2018LearningAnalyticsLowCost,Berthelot2025}. Advances in smaller, locally deployable LLMs make this shift increasingly feasible \cite{Liao2025Deepseek}. However, they introduce critical design decisions: prompt design, the crafting of instructions that guide LLMs to produce accurate and relevant outputs; batching strategies, which group multiple inputs for simultaneous processing rather than handling them individually; and model configurations, referring to prompt design and batch settings, all of which can substantially influence coding accuracy, processing time, and energy consumption. These factors create trade-offs that must be purposely engineered rather than assumed \cite{SamaraweeraEtAl2026LLMHealthcareSimulation}.

In this work, we examine how LLM prompting designs, including batching strategies that aggregate multiple inference requests into a single run, can be optimised to balance coding performance, processing time, and environmental impact in high-stakes, time-dependent contexts, such as team-based healthcare simulation. 

Healthcare simulations are widely used in medical and nursing education, as well as in interdisciplinary hospital training, to nurture communication and clinical skills in complex clinical scenarios such as patient deterioration, emergency response, and interprofessional handovers \cite{Fanning2007,Rosen2012}. These simulations require teams to coordinate, share critical information, anticipate risks, and make rapid decisions—processes that are enacted and observable through dialogue \cite{Sarangi2017}. Decades of research in healthcare teamwork show that communication quality is directly associated with improved patient safety and clinical outcomes \cite{Sun2018TeamworkSurgery}. The ability to identify and foster effective teamwork behaviours, such as closed-loop communication, information sharing, clarification, and mutual monitoring, is therefore of both educational and clinical significance \cite{zhao2023METS}.

A central pedagogical component of healthcare simulations is the debriefing phase, during which participants reflect on their teamwork and decision-making processes \cite{Abulebda2022}. Automated dialogue coding has the potential to render patterns of communication visible during or immediately after simulation, providing structured evidence to support reflection and targeted feedback \cite{Zhao2024endusers}. However, these contexts pose distinctive constraints. Simulations are often conducted in secure university facilities or hospital environments where cloud-based services may be restricted due to connectivity limitations, data governance policies, or ethical considerations regarding clinical scenarios \cite{SamaraweeraEtAl2026LLMHealthcareSimulation,Stenseth2025,Algarni2025}. Therefore, dialogue-coding tools must operate efficiently under constrained (often on-premises) deployment and produce outputs within the tight temporal boundaries of the debriefing workflow.

While prior work has explored automated dialogue coding in simulation settings using NLP models such as BERT \cite{zhao2023METS,wang2024artificial}, these approaches can be difficult to scale, adapt, and transfer across varying clinical scenarios and institutional contexts. The emergence of LLMs offers greater flexibility, but raises new questions about how to balance coding performance with computational efficiency and environmental impact in authentic deployment conditions. Addressing this gap, we systematically investigated how prompt design and batching strategies affect accuracy, processing time, and energy consumption when coding $11,647$ utterances across six team dialogue constructs. By explicitly examining these trade-offs, we move beyond feasibility toward actionable design guidance for deploying LLM-based dialogue analytics in high-stakes educational environments, where timely feedback, privacy, and computational and environmental impacts are critical.

\section{Background and Related Work}

\subsection{Automated Coding of Dialogue}
To support qualitative coding at scale, automated analysis has become increasingly important \cite{wang2024artificial}. Early efforts relied on traditional NLP approaches, including supervised classifiers and rule-based systems \cite{crowston2012using}. Although these methods demonstrated that dialogue could be computationally classified, they required substantial labelled training data and were tightly coupled to specific datasets and coding schemes, limiting transferability across contexts and theoretical constructs in CSCL.

Pre-trained transformer models, such as BERT, advanced this work by reducing dependence on large, task-specific annotated datasets (e.g., \cite{zhao2023METS,Zhao2024endusers,devlin2019BERT}). Fine-tuning improved dialogue coding compared to earlier machine learning approaches. Yet, these models still require task-specific annotated data and often struggle with the nuance, ambiguity, and interpretive flexibility inherent in qualitative coding--particularly when codes represent complex collaborative processes or context-dependent constructs \cite{Zhao2024BJET,zhao2023METS}.

LLMs extend these capabilities through large-scale pre-training and inference-based modelling. Using zero-shot or few-shot prompting, they can perform qualitative coding with minimal additional labelled data, making them practical for contexts where annotated corpora are scarce \cite{garg2024automated}. Recent studies illustrate this shift. Liu et al., \cite{Liu2025Qualitative} showed that GPT-4 can approximate human coding across prompting designs, though effectiveness varies by construct and context. Wang et al., \cite{Wang2025BERTvsLLM} found that LLaMA, while less precise than BERT, still produced pedagogically meaningful insights for teachers analysing dialogic practice. In large-scale MOOC settings, Mehta et al., \cite{Mehta2025} demonstrated strong agreement between GPT-assisted coding and human annotations when linking discussion engagement to learning outcomes. Similarly, Li et al., \cite{Li2025SSRL} applied GPT-4 and DeepSeek to identify phases of social shared regulation in online discussions, achieving over 83\% accuracy and illustrating how LLMs can operationalise theoretically rich CSCL constructs at scale. These studies prove that LLMs can generate contextually appropriate outputs for qualitative dialogue coding tasks~\citep{Li2025SSRL, Liu2025Qualitative}.

Despite these advances, evaluation of LLM outputs has focused predominantly on agreement with human coders, positioning accuracy as the primary success indicator. Yet deployment in authentic educational settings requires broader performance considerations \cite{Martinez2017}. First, processing time is critical~\cite{Cheng2023,Ji2025} as latency determines whether coded outputs can support timely reflection or intervention. Second, coding performance must be examined in relation to prompt and configuration choices, rather than treated as a static property of the model. Third, prompting designs—including zero-shot and few-shot approaches—shape computational load and environmental impact (CO\textsubscript{2}e), raising sustainability considerations for large-scale deployment \cite{Inie2025CO2STLY}. Yet, despite acknowledged contextual variability in coding effectiveness \cite{Liu2025Qualitative}, limited attention has been paid to how prompting configurations jointly affect processing time, coding performance, and environmental impact.

These issues are particularly salient in healthcare education, where dialogue unfolds under time pressure and outputs may inform near real-time debriefing \cite{Dieckmann2009}. To our knowledge, only two studies have applied LLM-based coding within authentic healthcare simulation. Samaraweera et al. \cite{SamaraweeraEtAl2026LLMHealthcareSimulation} examined prompting techniques for a hospital-based simulation operating under strict time constraints, while Tscholl et al. \cite{Tscholl2026} used AI-generated teamwork reports to support debriefing via thematic analysis. Though both demonstrated feasibility, neither systematically examined how prompting and batching configurations jointly influence processing time, coding performance, and environmental impact. %Addressing this gap requires moving beyond accuracy toward an integrated examination of LLM-based dialogue coding under the temporal, operational, and sustainability constraints of authentic healthcare simulations.

\subsection{Energy and Efficiency Trade-offs in LLM Inference}

LLM inference refers to the serving phase in which a trained model generates outputs in response to input prompts. 
In educational contexts, inference enables scalable applications such as automated encoding of communication among learners and educators \cite{Jin2024optimising,Mehta2025}, feedback generation \cite{BArno2024feedback,Nguyen2024comp}, adaptive scaffolding \cite{Nie2025,Zhang2024cflow}, and the creation of instructional or assessment materials \cite{Hutt2024scaling,Moore2024automated}. In this study, coding team dialogue is itself an inference task: the model is repeatedly executed to classify utterances in near real-time. As these systems transition from controlled settings into authentic learning contexts, inference, rather than training, becomes the dominant and recurring source of computational demand.

The distinction between model training and model inference matters. Although model training is energy-intensive, recent studies show that the energy consumption of LLM serving has now surpassed that of training, contributing substantially to carbon footprints measured in CO2e \cite{Chien2023reducing,ding2024sustainable}. Ding and Shi \cite{ding2024sustainable} further emphasised the limited understanding of trade-offs between performance and carbon in LLM serving, noting that efficiency needs, hardware configurations, and energy use are tightly interdependent. For dialogue analytics deployed at scale, where inference is executed continuously across sessions and cohorts, processing time and environmental impact are not peripheral concerns but key design constraints.

Moreover, constraints related to processing time and environmental impact during LLM inference are amplified in face-to-face authentic educational environments, where local LLM deployment (e.g., \cite{Liao2025Deepseek}) is often necessary due to privacy, governance, or connectivity limitations \cite{SamaraweeraEtAl2026LLMHealthcareSimulation,Martinez-Maldonado2023tochi}. Parameter-efficient approaches show that LLMs can be adapted for classroom dialogue analysis with reduced fine-tuning overhead \cite{Wang2025LORA}, yet inference itself remains computationally demanding. In multimodal learning analytics scenarios, where dialogue analysis may operate alongside additional data streams (e.g., indoor position or heart rate data), inefficient inference can introduce latency that undermines timely feedback \cite{SamaraweeraEtAl2026LLMHealthcareSimulation,Yan2025}, particularly during simulation debriefing, where reflection depends on immediacy \cite{Abulebda2022,Dieckmann2009}. Accordingly, analysing dialogue in authentic educational environments requires more than acceptable accuracy; it demands deliberate optimisation of efficiency and environmental impact to enable feasible, ethical, and scalable deployment.

\subsection{Contribution and Research Questions}

Situated within CSCL and team learning research, this study advances automated dialogue analysis by reframing performance beyond accuracy and examining how LLM-based coding can be design for authentic, high-stakes learning environments. Specifically, we investigate how prompt design and batching strategies shape the trade-offs between processing time, coding performance, and environmental impact in healthcare simulation, where timely feedback, local deployment, and sustainability are integral to supporting reflective practice and effective teamwork. By aligning technical optimisation with pedagogical goals, this work contributes design-oriented insights for scaling dialogue analytics in face-to-face collaborative learning contexts. Motivated by these considerations, we pose the following research questions:
\begin{itemize}

\item \textbf{RQ1:} How do prompt design and batching strategies affect processing time when using LLMs for team dialogue coding in healthcare simulation?
    % TODO Check the RQ1 to be more aligned affect rather than improved. Balancing Coding Performance, Processing Time, and Environmental Impact
\item \textbf{RQ2:} How do prompt design and batching strategies affect coding performance (accuracy) when using LLMs for team dialogue coding in healthcare simulation?

\item \textbf{RQ3:} To what extent do prompt design and batching strategies influence the environmental impact of LLM inference for team dialogue coding in healthcare simulation?

\end{itemize}

We pose a research question that integrates these dimensions:

\begin{itemize}
\item \textbf{RQ4:} Which prompt design and batching strategies enable a favourable trade-off between coding performance (accuracy), processing time, and environmental impact for team dialogue coding in healthcare simulation?
\end{itemize}

\section{Method}

%To address the research questions, we first set up a locally deployed large language model (LLM) using Ollama\footnote{https://www.ollama.com}. We then examined four prompt designs that differed in the number of simultaneous utterances (batch size) used for qualitative coding. The utterances were drawn from a previously collected simulation dataset. We evaluated the impact of the prompt designs and batch size on coding performance, processing time, and environmental impact.

%Finally, to address RQ4, we jointly analysed coding performance, processing time, and energy consumption and conducted a trade-off analysis to identify configurations that balance these competing objectives.

\subsection{Learning Context and Learning Design}
In nursing education, healthcare simulations are used to replicate emergency scenarios in hospital wards. Nursing learners practise teamwork, coordination, and clinical communication within this face-to-face high-fidelity setting.
%The simulations took place over a four-week period between August and September in $2021$ and $2022$. During the simulations, audio data were collected from $140$ consenting learners, aged $20$ to $23$, organised into $35$ teams of four members.
The simulation scenario involves providing and prioritising care to four patients (i.e., manikins), recognising patient deterioration, calling for help, and managing the deterioration. As specified in the learning design, two learners, role-playing as primary nurses, initially participated in the scenario. Two additional learners, role-playing secondary nurses, entered after one primary nurse called for help. The simulation concluded once the deterioration was managed and the situation stabilised. %As the simulation progressed, the physical condition of one manikin patient deteriorated. learners were required to prioritise care for this deteriorating patient, which constituted their main task, and to declare an emergency state for this patient.

%After the emergency was declared, the other two learners, role-playing as secondary nurses, entered the simulation to assist. The four learners then collaborated to care for both the deteriorating patient and the remaining patients simultaneously. A doctor, enacted by a educator, also entered the room to diagnose the deteriorating patient, and learners were expected to communicate effectively with the doctor. The simulation concluded once the patient's deterioration issue was resolved.
A structured debriefing session follows each scenario to guide learners’ reflection on their teamwork and clinical decision-making. While the debriefing phase highlights the pedagogical need for timely, actionable dialogue analytics, the present study focuses exclusively on team dialogue collected during the scenario itself. The analysis of debriefing conversations falls beyond the scope of this paper. Thus, this learning setting reflects realistic constraints related to time, resources, and staff availability, commonly encountered in simulation-based healthcare education. These operational conditions frame the design considerations examined in this study.

\subsection{Dataset}

With ethics approval from Monash University Human Research Ethics Committee, all students provided consent for their data to be collected and analysed for this research. Students, aged $20$ to $23$, were organised into teams of four members. The dataset is formed from audio transcripts of $35$ teams: $15$ in $2021$ and $25$ in $2022$. 

Audio from each team member was recorded using Shure PGA31 headset microphones connected via BLX/SVX~4 wireless transmitters to a TASCAM US-16x08 audio interface, with all channels captured on a local server. Utterance timing for each student was extracted using a Python script and
a voice activity detection library \footnote{webrtcVAD: https://github.com/wiseman/py-webrtcvad}. Transcriptions were automatically generated with the Whisper-large model \cite{Radford2023}. $811$ utterances were manually transcribed (total duration = $5,078.01$ seconds) to assess the quality of the transcription. Based on these manual transcriptions, the word error rate (WER), a widely accepted metric for transcription quality \cite{Radford2023, Southwell2022}, of automatic transcriptions was 0.089. This result indicated the transcription quality was sufficient \cite{Southwell2022}.

The resulting transcripts from the $35$ teams were segmented into $11,647$ utterances. The average number of utterances per simulation was $332.8$ with a standard deviation of $93.9$. Utterances were broken within each dialogue segment into multiple sentences using punctuation marks, i.e., including full stops, commas, question marks, and exclamation marks. For the dialogue analysis, we used the coding scheme described in Table~\ref{tab:team_dialogue_coding}, which was derived from team science and collaboration frameworks \cite{Riley2008,Miller2009}. These codes were further refined with healthcare educators \cite{Zhao2024endusers}. The detailed theoretical grounding of the coding scheme is provided in \cite{Zhao2024BJET}.

The utterances were then coded at the sentence level, with each sentence assigned to a single code, thus multiple codes can appear within a single utterance. Two researchers independently coded $20\%$ of utterances collected in $2021$ ($756$ utterances from $3$ simulation sessions), using the definitions and examples provided in Table~\ref{tab:team_dialogue_coding}.
Cohen’s Kappa was used as the metric to assess the inter-rater reliability~\cite{McHugh2012Kappa}. A threshold of $0.61$ was chosen to indicate a substantial agreement between the raters~\cite{Landis1977}. For each code, a Kappa greater than $0.7$ was reached. Then, one researcher coded the remaining utterances in $2021$
and $2022$ datasets.

\begin{table}[t]
\centering
\caption{Coding used for team dialogue analysis, with definitions and examples.}
\label{tab:team_dialogue_coding}
\small
\setlength{\tabcolsep}{6pt}
    \begin{tabularx}{\linewidth}{l X}
        \toprule
        \textbf{Coding Scheme} & \textbf{Definition and Example} \\
        \midrule
        task allocation &
        A healthcare student self-assigns or assigns to another healthcare student one or more tasks. 
        \newline Example: "Can you start the IV line and prepare the fluids?" \\

        handover & 
        When a healthcare student provides structural communication of the medical situation to the new team members.
        \newline Example: "We have Jack here. He's just been complaining of six out of ten chest pain. " \\
        
        sharing information &
        A healthcare student shares relevant clinical or situational information with one or more other healthcare professionals.
        \newline Example: "His oxygen saturation has dropped to 88\%." \\
        
        escalation &
        A healthcare student communicates that the situation exceeds their capacity and requests or suggests additional assistance, including initiating or proposing a formal call for help.
        \newline Example: "This is deteriorating quickly, we need to call the resuscitation team." \\
        
        questioning &
        A healthcare student asks another team member to obtain information.
        \newline Example: "Have the blood results come back yet?" \\
        
        acknowledging &
        A healthcare student signals receipt or recognition of another professional's statement, request, or presence. This is a passive action and does not necessarily indicate agreement or disagreement, but demonstrates awareness or understanding.
        \newline Example: "Okay," or "I hear you." \\
        \bottomrule
    \end{tabularx}
\end{table}

Using the $11,647$ coded utterances, we filtered out those that were not assigned any communication code because they were either outside the context of the simulation (e.g., prior to or after the simulation) or contained no relevant information for the simulation task. The final dataset comprised $4,057$ utterances, which were used to address the research questions.

\subsection{Experimental Design}
For the experimental design, we created a set of prompt designs and batching strategies\footnote{Prompt and batching details can be found in  \\ https://osf.io/gtx86/overview?view\_only=ffdb65822459434e8f9344d8a263fb5a}. % TODO add the appendix using OSF links.

\subsubsection{Prompt Design and Batching Strategies}

We created four prompt designs following a few-shot classification approach~(i.e., Liu et al.~\cite{Liu2025Qualitative}). %The few-shot prompt included explicit code definitions and multiple annotated examples to guide model behaviour. 
\textbf{P.1: Prompt few-shot}, consisted of a contextual description specifying the model's role, definitions of the six communication codes plus three annotated examples per code, and task instructions framing the problem as a multi-label classification task. Each code was represented as a one-hot vector (i.e., $1$ indicating the presence of a label). The model's output consisted of seven binary values ($0$ or $1$) and was ordered according to the predefined codes.

\textbf{P.2: Prompt few-shot with rules}, extended the first by introducing explicit decision rules to support the coding of more complex dialogue codes~\cite{White2023,Sivarajkumar2024}. In addition to code definitions and examples, this prompt included a set of decision rules. For example: \emph{If an utterance contains a first-person commitment (e.g., "I will...", "Let me..."), then code it as} \texttt{task  allocation} \emph{; if the utterance also reports patient status, additionally code it as} \texttt{sharing information}.

\textbf{P.3: Prompt few-shot with rules and context}, incorporated preceding utterances as conversational context, given that modelling dialogue history has been shown to enhance robustness in conversational tasks~\cite{Gekhman2023}. Specifically, this design added a confidence-based rule (\emph{If uncertainty is below $95\%$, leave labels as $0$}) and modified the task instructions to require the model to consider the three preceding utterances when coding each target utterance.

\textbf{P.4: Prompt few-shot with rules and structured metadata}, extended P.2 by explicitly including contextual metadata, such as the initiator and receiver of each utterance (i.e., primary nurse 1, secondary nurse 1), to provide additional conversational structure.

In addition to prompt design, the current study investigated how batching strategies varied with the number of utterances processed simultaneously. Batching strategies are often used to analyse computational efficiency; for LLM inference, grouping multiple inputs into a single batch reduces repeated prompt parsing and token efficiency and can lower total inference time, thereby increasing scalability~\cite{Cheng2023,Ji2025}. Batch sizes of $1$, $10$, $20$, $30$, $40$, $50$, $60$, and $70$ utterances were evaluated. These batch sizes were heuristically selected to systematically explore the effect of increasing the number of utterances processed per batch in increments of ten. In total, this resulted in $32$ configurations (4 prompt designs $\times$ 8 batch sizes).

\subsection{LLM Implementation}
A local deployment of an LLM was configured. All model inference was conducted using the Ollama\footnote{https://www.ollama.com} interface with the Deepseek-R1\cite{Guo2025} model, more specifically the distilled model with 14 Billion parameters, \texttt{deepseek-r1:14b}~\footnote{https://ollama.com/library/deepseek-r1:14b}. Ollama was used to facilitate model deployment and provide an application interface (API) for accessing the DeepSeek model. The model temperature was set to $0$, resulting in deterministic outputs; therefore, repeated stochastic sampling was not required. To verify deterministic consistency across runs, each configuration was executed five times with batches of $20$ utterances, yielding identical outputs in all cases. To further support reproducibility, a fixed random seed of $42$ was used.

All evaluations were conducted on the same machine, a server equipped with $128$GB of $4,400$MHz ECC DDR5 RAM, an NVIDIA RTX A5500 GPU with $24$GB of memory, and two Intel Xeon Gold 5415+ processors ($2.9$GHz base frequency, $8$ cores each, $22.5$MB L3 cache, hyper-threading enabled; maximum turbo frequency $4.1$GHz). We record the timestamps for each of the experimental configurations. Running all required approximately $54$ hours and $41$ minutes of wall-clock time.

\subsection{Analysis}

To address \textbf{RQ1--RQ3}, we analysed $4,057$ utterances together with the researcher-coded ground truth to evaluate coding performance, processing time, and environmental impact. 

\subsubsection{RQ1 -- Processing Time}

We measured the total processing time required to code the complete set of $4,057$ utterances under each configuration. Processing time was analysed by estimating the time required to code the utterances from a single simulation session. Specifically, the total elapsed time was divided by the number of simulations in the dataset ($n=35$). We analysed the total elapsed time across the $32$ configurations (4 prompts $\times$ 8 batch sizes) and conducted Spearman’s rank correlation analyses to identify relationships among batch size, prompt design, and per-session processing time. Additionally, we visually inspected changes in processing time across prompt designs and batch sizes to identify configurations that support feasible timely deployments.

\subsubsection{RQ2 -- Coding Performance}
We evaluated coding performance using standard metrics, including macro-averaged F1 score, precision, recall, and accuracy \cite[e.g.,][]{Liu2025Qualitative}. Performance was computed at the utterance level, resulting in a multi-label classification evaluation. F1 score was used as the primary performance metric due to its robustness to class imbalance in our dataset (see Appendix). These metrics were used to analyse the relationship between batch size, prompt design, and performance, and to determine whether variations in batch size and prompt design affected F1 scores.

\subsubsection{RQ3 -- Environmental Impact}
We measured the total energy consumed while processing the full dataset for each configuration. Energy measurements included CPU, GPU, and DRAM usage, which were aggregated to obtain total system energy consumption. Energy usage was recorded using the Zeus Python library~\cite{You2023} and reported in joules\footnote{A joule is an International System unit of energy, equal to applying a force of one newton over one metre, or roughly the energy to lift a $100$ apple by 1 metre.}. Rather than estimating environmental impact using carbon dioxide equivalent (CO$_{2}$e) calculators, we report energy consumption directly as a proxy for environmental impact. This choice is motivated by the fact that carbon and water consumption calculators are derived from energy usage~\cite{Elsworth2025}. Furthermore, all experiments were conducted on local deployments without active water-based cooling.

\subsubsection{RQ4 -- Trade-offs}
We conducted a trade-off analysis based on Pareto optimality~\cite{Ngatchou2005,stadler1988multicriteria}. Pareto analysis is a widely used approach for evaluating trade-offs among conflicting objectives, where improving one objective may degrade others. It provides a systematic method for identifying configurations that achieve optimal trade-offs across multiple criteria.

This analysis considered three objectives: (i) coding performance, measured by the macro-averaged F1 score; (ii) total GPU energy consumption; and (iii) total processing time. Before identifying Pareto-optimal configurations, we conducted a Spearman’s rank correlation analysis between GPU energy consumption and total processing time to examine the relationship between these dimensions. Pareto-optimal configurations were identified by maximising the F1 score while simultaneously minimising energy consumption and processing time. The resulting Pareto-optimal set comprises non-dominated configurations, for which no alternative configuration achieves better performance on one objective without degrading at least one of the others. This set is commonly referred to as the Pareto front~\cite{Ngatchou2005}. We filtered out configurations infeasible for our near real-time scenario. We further analysed per-label results within the Pareto front to identify configurations that provide a balanced trade-off across all three objectives.

\section{Results}

\subsection{RQ1 -- Processing Time}

\begin{figure*}[htpb]
    \centering
    % \includegraphics[width=0.75\linewidth]{figures/time_vs_batch.pdf}
    % \caption{Processing Time in seconds and Batch sizes.}

    \begin{subfigure}{0.45\linewidth}
        \centering
        \includegraphics[width=0.85\linewidth]{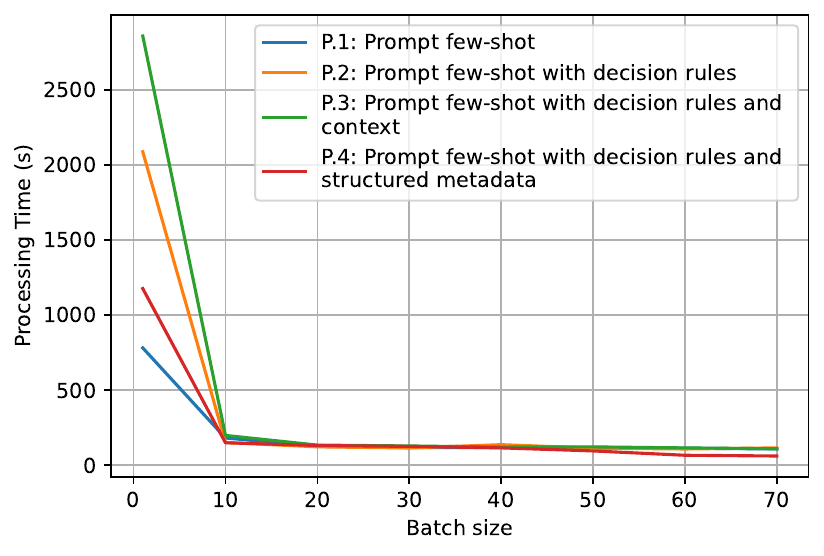}
        \caption{Processing Time against Batch sizes.}
        \label{fig:processing_time_by_prompt}
    \end{subfigure}
    \hfill
    \begin{subfigure}{0.45\linewidth}
        \centering
        \includegraphics[width=0.85\linewidth]{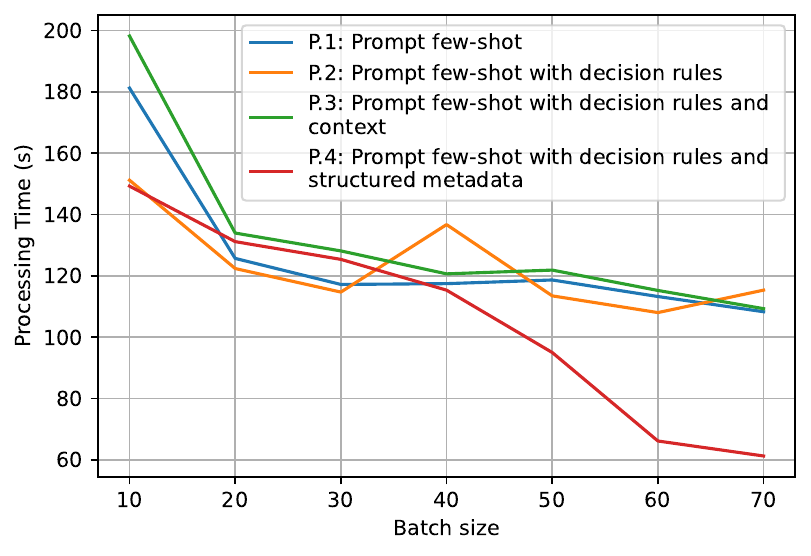}
        \caption{Processing Time against Batch sizes for Batch $\geq 10$}
        \label{fig:processing_time_by_prompt_configuration_zoom}
    \end{subfigure}
    \caption{Processing Time in Seconds across prompts designs and batch sizes.}
    \label{fig:processing_time_by_prompt_configuration}
    \Description{This figure contains two subfigures. The first shows processing time across batch sizes starting from $1$, while the second shows results starting from a batch size of $10$.}
\end{figure*}

The processing times (seconds) required to code a single simulation session are presented in Table~\ref{tab:processing_time_by_prompt_configuration_tab}. The results for the entire set of utterances are provided in the Appendix. Table~\ref{tab:processing_time_by_prompt_configuration_tab} and Fig.~\ref{fig:processing_time_by_prompt} shows a clear trend, as batch size increases, total processing time decreases across all prompt designs. 
Notably, for \emph{P.1: Prompt few-shot}, a batch size of $1$ required approximately four times more processing time than a batch size of $10$. For \emph{P4: Prompt few-shot with rules and structured metadata}, this difference increased to approximately ten times. Similarly, \emph{P.2: Prompt few-shot with rules} and \emph{P3: Prompt few-shot with rules and context} required approximately fourteen times more processing time at batch size $1$ compared to batch size $10$.

As shown in Fig.~\ref{fig:processing_time_by_prompt_configuration_zoom}, processing time began to stabilise beyond a batch size of $20$ for most prompt designs, with the exception of \emph{P4}. At batch size $70$, \emph{P4} achieved the lowest processing time, requiring approximately half the time ($0.55 \times$) compared to the other prompts. For the remaining prompts, increasing the batch size beyond $20$ results in relatively small reductions, ranging from $3$ to $8$ seconds.
Spearman's rank correlation between batch size and processing time was $-0.895$, indicating a strong negative association: larger batch sizes were consistently linked to shorter processing times.
In contrast, configurations with batch size $1$ (i.e., no batching) were impractical for near real-time deployment, as they required more than $13$ minutes per session. %This is impractical for our learning situation, as the debriefing starts immediately after the simulation ends.

Overall, these findings suggest that processing time was substantially affected by batching strategies. In this context, no batching is impractical because the debriefing starts immediately after the simulation ends. A batch size of 20-70 may be optimal at this point.

\begin{table}[b]
    \caption{Processing Time in seconds across prompt designs and batch sizes. Highlighted values indicate the minimum value per prompt and metric, while values in bold denote the overall maximum for each metric. P.1: Prompt few-shot, P.2: Prompt few-shot with rules, P.3: Prompt few-shot with rules and context, P.4: Prompt few-shot with rules and structured metadata. }
    \label{tab:processing_time_by_prompt_configuration_tab}
    \begin{tabularx}{\linewidth}{c *{4}R}
    \toprule
    \multirow{2}{*}{\textbf{Batch}} & \multicolumn{4}{c}{\textbf{Prompt}} \\
    \cmidrule(lr){2-5}
    & {\textbf{P.1}} & {\textbf{P.2}} & {\textbf{P.3}} & {\textbf{P.4}} \\
    \midrule
    1  & 781.03  & 2,086.90 & 2,857.92 & 1,176.21 \\
    10 & 181.23  & 151.20  & 198.20  & 149.26 \\
    20 & 125.69  & 122.39  & 133.96  & 131.16 \\
    30 & 117.20  & 114.72  & 128.14  & 125.38 \\
    40 & 117.46  & 136.72  & 120.66  & 115.35 \\
    50 & 118.66  & 113.49  & 121.89  & 95.01  \\
    60 & 113.27  & \cellcolor{Gray}108.03  & 115.25  & 66.17  \\
    70 & \cellcolor{Gray}108.31  & 115.35  & \cellcolor{Gray}109.31  & \cellcolor{Gray}\textbf{61.28}  \\
    \bottomrule

    \end{tabularx}
\end{table}

\subsection{RQ2 -- Coding Performance}
\begin{figure}[t]
    \centering
    \includegraphics[width=0.85\linewidth]{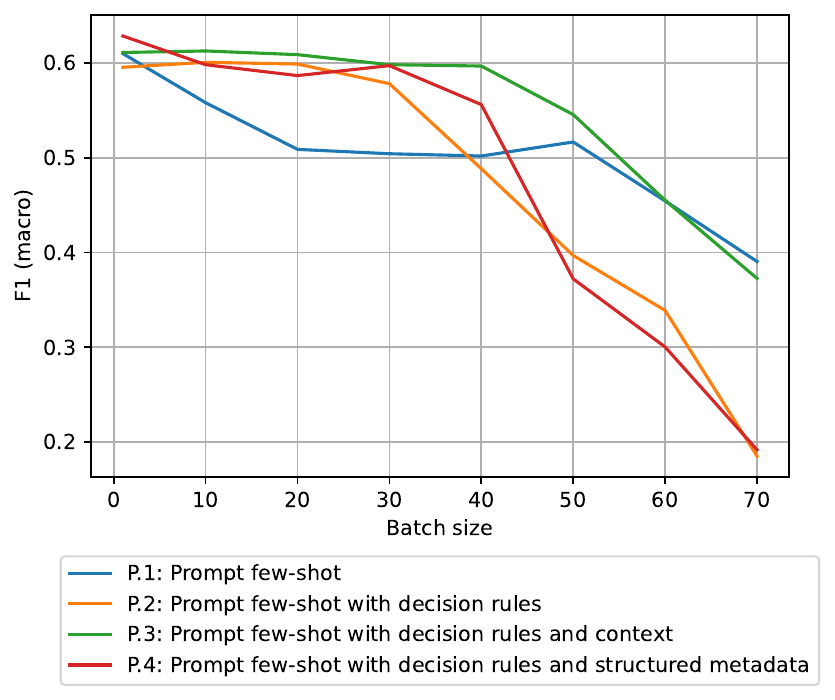}
    \caption{Macro-averaged F1 scores and batch sizes illustrating coding performance.}
    \label{fig:f1macro_vs_batch}
    \Description{This figure shows F1 scores across batch sizes from $1$ to $70$. Each line represents the performance of a different prompt design.}
\end{figure}

%To address \textbf{RQ2}, we evaluated the impact of prompt design and batch size on coding performance using macro-averaged F1 score, precision, and recall. 
The results evaluating coding performance when considering prompt design and batch size (macro-averaged F1 score, precision, and recall) are reported in Table~\ref{tab:performance_by_prompt_configuration}. %Highlighted values indicate the maximum value per prompt for each metric, while bolded values denote the overall maximum per metric. These highlights 
These results indicate that higher performance was generally achieved at lower batch sizes. As shown in Fig.~\ref{fig:f1macro_vs_batch}, there was a strong negative association between batch size and F1 score, with Spearman's rank correlation of $-0.882$, indicating that performance consistently declined as batch size increased, suggesting that higher batch sizes did not necessarily improve the model's accuracy.

\begin{table}[b]
\centering
\caption{Coding performance across prompt designs. Highlighted values indicate the maximum value per prompt for each metric, while bolded values denote the overall maximum for each metric.}
\label{tab:performance_by_prompt_configuration}
\small
\setlength{\tabcolsep}{1pt}
\begin{tabularx}{\linewidth}{c *{12}{>{\centering\arraybackslash}X}}
\toprule
\multirow{2}{*}{\textbf{Batch}} & \multicolumn{4}{c}{\textbf{F1 Macro}} 
& \multicolumn{4}{c}{\textbf{Precision Macro}} 
& \multicolumn{4}{c}{\textbf{Recall Macro}} \\
\cmidrule(lr){2-5} \cmidrule(lr){6-9} \cmidrule(lr){10-13}
& \textbf{P.1} & \textbf{P.2} & \textbf{P.3} & \textbf{P.4} & \textbf{P.1} & \textbf{P.2} & \textbf{P.3} & \textbf{P.4} & \textbf{P.1} & \textbf{P.2} & \textbf{P.3} & \textbf{P.4} \\
\midrule
1  & \cellcolor{Gray}0.61 & \cellcolor{Gray}0.60 & \cellcolor{Gray}0.61 & \cellcolor{Gray}\textbf{0.63} & 0.63 & 0.67 & 0.67 & \cellcolor{Gray}\textbf{0.68} & \cellcolor{Gray}0.61 & \cellcolor{Gray}0.59 & 0.61 & \cellcolor{Gray}\textbf{0.63} \\
        10 & 0.56 & \cellcolor{Gray}0.60 & \cellcolor{Gray}0.61 & 0.60 & \cellcolor{Gray}0.64 & \cellcolor{Gray}\textbf{0.68} & \cellcolor{Gray}\textbf{0.68} & \cellcolor{Gray}\textbf{0.68} & 0.52 & \cellcolor{Gray}0.59 & \cellcolor{Gray}0.62 & 0.57 \\
        20 & 0.51 & \cellcolor{Gray}0.60 & \cellcolor{Gray}0.61 & 0.59 & 0.61 & \cellcolor{Gray}\textbf{0.68} & 0.67 & 0.66 & 0.45 & 0.58 & 0.60 & 0.55 \\
        30 & 0.50 & 0.58 & 0.60 & 0.60 & 0.63 & 0.66 & 0.64 & 0.67 & 0.43 & 0.54 & 0.59 & 0.56 \\
        40 & 0.50 & 0.49 & 0.60 & 0.56 & 0.63 & 0.65 & 0.65 & 0.65 & 0.42 & 0.42 & 0.58 & 0.50 \\
        50 & 0.52 & 0.40 & 0.55 & 0.37 & 0.63 & 0.54 & 0.64 & 0.53 & 0.46 & 0.33 & 0.50 & 0.30 \\
        60 & 0.45 & 0.34 & 0.46 & 0.30 & 0.58 & 0.59 & 0.65 & 0.45 & 0.39 & 0.25 & 0.37 & 0.24 \\
        70 & 0.39 & 0.19 & 0.37 & 0.19 & 0.55 & 0.51 & 0.61 & 0.33 & 0.32 & 0.12 & 0.28 & 0.14 \\
\bottomrule
\end{tabularx}
\end{table}

Regarding prompt strategies, configurations with explicit decision rules consistently outperformed the \emph{Prompt few-shot} baseline. This finding suggests that structured decision rules enhanced the model’s ability to differentiate dialogue codes, particularly at batch sizes below $40$.

Given the multi-label nature of the task, we further report per-label F1 scores in Tables~\ref{tab:per_label_ack_escalation_handover} and \ref{tab:per_label_questioning_sharing_allocation}. At the code level, batch size continued to influence performance. Prompts that included decision rules generally outperformed the \emph{P.1: Prompt few-shot} across most dialogue codes. In particular, \emph{P4: Prompt few-shot with rules and structured metadata} achieved the strongest performance in three of the six codes—\emph{acknowledging}, \emph{handover}, and \emph{sharing information}, although it showed greater instability at larger batch sizes.

Beyond peak performance values, as shown in Fig.~\ref{fig:f1macro_vs_batch}, we observed that prompts incorporating explicit decision rules (P.2, P.3 and P.4) demonstrate greater robustness to increasing batch sizes compared to the baseline few-shot prompt. While performance declined across all configurations as batch size increased, rule-based prompts exhibited slower degradation, particularly for \emph{acknowledging} and \emph{escalation} codes. Furthermore, \emph{handover} remained consistently the most challenging label across all prompt designs and batch sizes, with substantially lower F1 scores compared to other codes, suggesting difficulty in modelling this dialogue code. Notably, moderate batch sizes (10-20) often yielded better performance across multiple codes, suggesting that such sizes may support contextual grouping before attention dilution affects classification accuracy~\cite{Zhu2024}.

In sum, these findings suggest that coding performance, as measured by F1 score, is strongly influenced by batching strategies, with smaller batch sizes ($n = 1$-$10$) consistently supporting more accurate classification.

\begin{table}[ht]
\centering
\caption{Per-label F1 scores across prompt designs and batch sizes.}
\label{tab:per_label_ack_escalation_handover}
\small
\setlength{\tabcolsep}{4pt}
\begin{tabularx}{\linewidth}{c *{12}{>{\centering\arraybackslash}X}}
\toprule
        \multirow{2}{*}{\textbf{Batch}} & \multicolumn{4}{c}{\textbf{acknowledging}} 
        & \multicolumn{4}{c}{\textbf{escalation}} 
        & \multicolumn{4}{c}{\textbf{handover}} \\
        \cmidrule(lr){2-5} \cmidrule(lr){6-9} \cmidrule(lr){10-13}
        & \textbf{P.1} & \textbf{P.2} & \textbf{P.3} & \textbf{P.4} & \textbf{P.1} & \textbf{P.2} & \textbf{P.3} & \textbf{P.4} & \textbf{P.1} & \textbf{P.2} & \textbf{P.3} & \textbf{P.4} \\
        \midrule
        1  & \cellcolor{Gray}0.73 & \cellcolor{Gray}0.78 & \cellcolor{Gray}0.79 & \cellcolor{Gray}\textbf{0.82} & \cellcolor{Gray}0.66 & 0.68 & 0.65 & \cellcolor{Gray}0.67 & 0.24 & 0.20 & 0.19 & 0.19 \\
        10 & 0.64 & 0.74 & 0.77 & 0.72 & 0.60 & 0.63 & 0.64 & 0.66 & \cellcolor{Gray}0.27 & 0.16 & 0.16 & 0.18 \\
        20 & 0.58 & 0.72 & 0.75 & 0.71 & 0.48 & \cellcolor{Gray}\textbf{0.70} & 0.65 & 0.63 & 0.23 & 0.15 & 0.17 & 0.19 \\
        30 & 0.58 & 0.70 & 0.73 & 0.70 & 0.50 & 0.55 & 0.64 & 0.67 & 0.24 & \cellcolor{Gray}0.22 & 0.18 & \cellcolor{Gray}\textbf{0.26} \\
        40 & 0.58 & 0.62 & 0.73 & 0.67 & 0.52 & 0.48 & \cellcolor{Gray}0.66 & 0.57 & 0.25 & 0.11 & \cellcolor{Gray}0.20 & 0.22 \\
        50 & 0.62 & 0.53 & 0.68 & 0.49 & 0.56 & 0.43 & 0.57 & 0.27 & 0.26 & 0.04 & 0.13 & 0.16 \\
        60 & 0.51 & 0.45 & 0.59 & 0.47 & 0.53 & 0.39 & 0.45 & 0.28 & 0.18 & 0.04 & \cellcolor{Gray}0.20 & 0.23 \\
        70 & 0.45 & 0.24 & 0.47 & 0.34 & 0.41 & 0.19 & 0.39 & 0.21 & 0.18 & 0.06 & 0.13 & 0.01 \\
        \bottomrule
\end{tabularx}
\end{table}

\begin{table}[ht]
\centering
\caption{Per-label F1 scores across prompt designs and batch sizes.}
\label{tab:per_label_questioning_sharing_allocation}
\footnotesize
\begin{tabularx}{\linewidth}{c *{12}{>{\centering\arraybackslash}X}}
\toprule
        \multirow{2}{*}{\textbf{Batch}} & \multicolumn{4}{c}{\textbf{questioning}} 
        & \multicolumn{4}{c}{\textbf{sharing info.}} 
        & \multicolumn{4}{c}{\textbf{task allocation}} \\
        \cmidrule(lr){2-5} \cmidrule(lr){6-9} \cmidrule(lr){10-13}
        & \textbf{P.1} & \textbf{P.2} & \textbf{P.3} & \textbf{P.4} & \textbf{P.1} & \textbf{P.2} & \textbf{P.3} & \textbf{P.4} & \textbf{P.1} & \textbf{P.2} & \textbf{P.3} & \textbf{P.4} \\
        \midrule
        1  & \cellcolor{Gray}\textbf{0.85} & 0.80 & 0.81 & \cellcolor{Gray}0.84 & \cellcolor{Gray}0.55 & 0.53 & 0.56 & \cellcolor{Gray}\textbf{0.59} & \cellcolor{Gray}0.63 & 0.58 & 0.67 & \cellcolor{Gray}0.66 \\
        10 & 0.76 & \cellcolor{Gray}0.82 & \cellcolor{Gray}0.83 & 0.81 & 0.52 & \cellcolor{Gray}0.57 & \cellcolor{Gray}0.58 & 0.57 & 0.57 & \cellcolor{Gray}\textbf{0.69} & \cellcolor{Gray}\textbf{0.69} & 0.65 \\
        20 & 0.71 & 0.80 & 0.82 & 0.79 & 0.51 & \cellcolor{Gray}0.57 & \cellcolor{Gray}0.58 & 0.56 & 0.54 & 0.66 & \cellcolor{Gray}\textbf{0.69} & 0.64 \\
        30 & 0.71 & 0.80 & 0.80 & 0.78 & 0.51 & \cellcolor{Gray}0.57 & \cellcolor{Gray}0.58 & 0.56 & 0.48 & 0.62 & 0.66 & 0.61 \\
        40 & 0.70 & 0.73 & 0.79 & 0.76 & 0.48 & 0.50 & 0.55 & 0.54 & 0.48 & 0.49 & 0.65 & 0.58 \\
        50 & 0.73 & 0.62 & 0.77 & 0.54 & 0.50 & 0.44 & 0.53 & 0.41 & 0.42 & 0.31 & 0.59 & 0.37 \\
        60 & 0.65 & 0.52 & 0.61 & 0.37 & 0.47 & 0.38 & 0.43 & 0.30 & 0.38 & 0.25 & 0.45 & 0.14 \\
        70 & 0.55 & 0.30 & 0.53 & 0.22 & 0.41 & 0.20 & 0.37 & 0.27 & 0.35 & 0.11 & 0.35 & 0.09 \\
        \bottomrule
\end{tabularx}
\end{table}

\begin{table}
\caption{GPU Energy consumption in Joules per simulation across prompt designs and batch sizes}
\label{tab:energy_consumption}
\small
\begin{tabularx}{\linewidth}{c *{4}R}
\toprule
\multirow{2}{*}{\textbf{Batch}} & \multicolumn{4}{c}{\textbf{Prompt}} \\
\cmidrule(lr){2-5}
& {\textbf{P.1}} & {\textbf{P.2}} & {\textbf{P.3}} & {\textbf{P.4}} \\
\midrule
$1$  & 177.46 & 474.85 & 652.46 & 267.85 \\
$10$ & 41.40  & 34.49  & 45.30  & 34.07 \\
$20$ & 28.71  & 27.97  & 30.62  & 30.00 \\
$30$ & 26.80  & 26.23  & 29.31  & 28.67 \\
$40$ & 26.87  & 31.28  & 27.60  & 26.39 \\
$50$ & 27.14  & 25.95  & 27.88  & 21.72 \\
$60$ & 25.90  & \cellcolor{Gray}24.71  & 26.37  & 15.12 \\
$70$ & \cellcolor{Gray}24.59  & 26.40  & \cellcolor{Gray}25.00  & \cellcolor{Gray}\textbf{14.01} \\
\bottomrule
\end{tabularx}
\end{table}

\begin{figure*}
    \centering
    % \includegraphics[width=0.75\linewidth]{figures/time_vs_batch.pdf}
    % \caption{Processing Time in seconds and Batch sizes.}

    \begin{subfigure}{0.45\linewidth}
        \centering
        \includegraphics[width=0.85\linewidth]{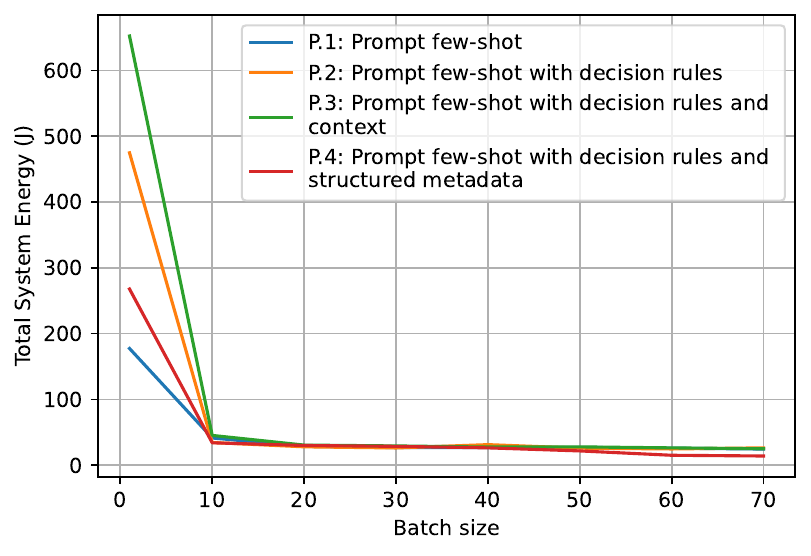}
        \caption{Energy Consumption against Batch sizes.}
    \end{subfigure}
    \hfill
    \begin{subfigure}{0.45\linewidth}
        \centering
        \includegraphics[width=0.85\linewidth]{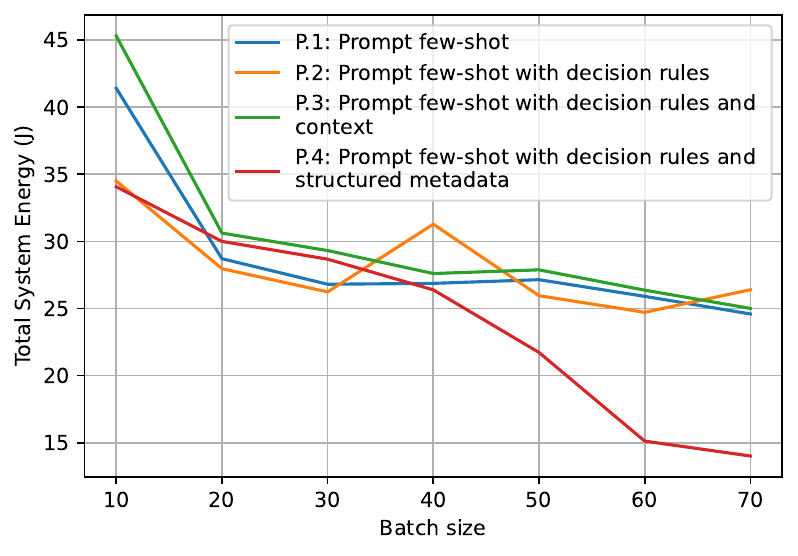}
        \caption{Energy Consumption against Batch sizes for Batch $\geq 10$}
    \end{subfigure}
    \label{fig:}
    \caption{Energy Consumption in Joules across prompts designs and batch sizes.}
    \label{fig:energy_consumption_by_prompt_configuration}
    \Description{This figure contains two subfigures. The first shows Energy consumption across batch sizes starting from $1$, while the second shows results starting from a batch size of $10$.}
\end{figure*}

\subsection{RQ3 -- Environmental Impact}

To address the environmental impact of prompt design and batching strategies, we analysed the effect of these configurations on energy consumption (Table~\ref{tab:energy_consumption}). As expected, batch size strongly influenced energy use, closely mirroring its relationship with processing time. In general, larger batch sizes led to lower total energy consumption due to reduced computational overhead per utterance.

Across prompt designs, the \emph{Prompt few-shot with rules and context} configuration exhibited the highest energy consumption at batch size $1$, consistent with its longer processing time. In contrast, the \emph{Prompt few-shot with rules} showed energy consumption comparable to the \emph{Prompt few-shot} baseline across most batch sizes. These findings indicate that incorporating explicit decision rules did not substantially increase energy costs when combined with appropriate batching strategies.

Comparing the results for Processing Time (Fig.~\ref{fig:processing_time_by_prompt_configuration}) and Energy Consumption (Fig.~\ref{fig:energy_consumption_by_prompt_configuration}), both showed highly similar patterns. Results from Spearman's rank correlation examining the relationship between energy consumption and processing time showed an almost perfect positive correlation (Spearman's rank correlation coefficient of 0.99945), confirming that energy consumption scaled nearly proportionally with processing time.

\subsection{RQ4 - Trade-offs}

% To address \textbf{RQ4}, we conducted a multi-objective trade-off analysis considering macro-averaged F1 score (coding performance), total processing time, and total system energy consumption. The objective was to identify non-dominated configurations that maximise performance while minimising computational cost and environmental impact.
Given the near-perfect correlation between processing time and energy consumption (Spearman's rank $0.99945$), we reduced the optimisation problem to two objectives: coding performance and processing time. In this context, minimising processing time effectively minimises energy consumption.

\begin{figure*}[ht]
    \centering
    \includegraphics[width=0.65\linewidth] {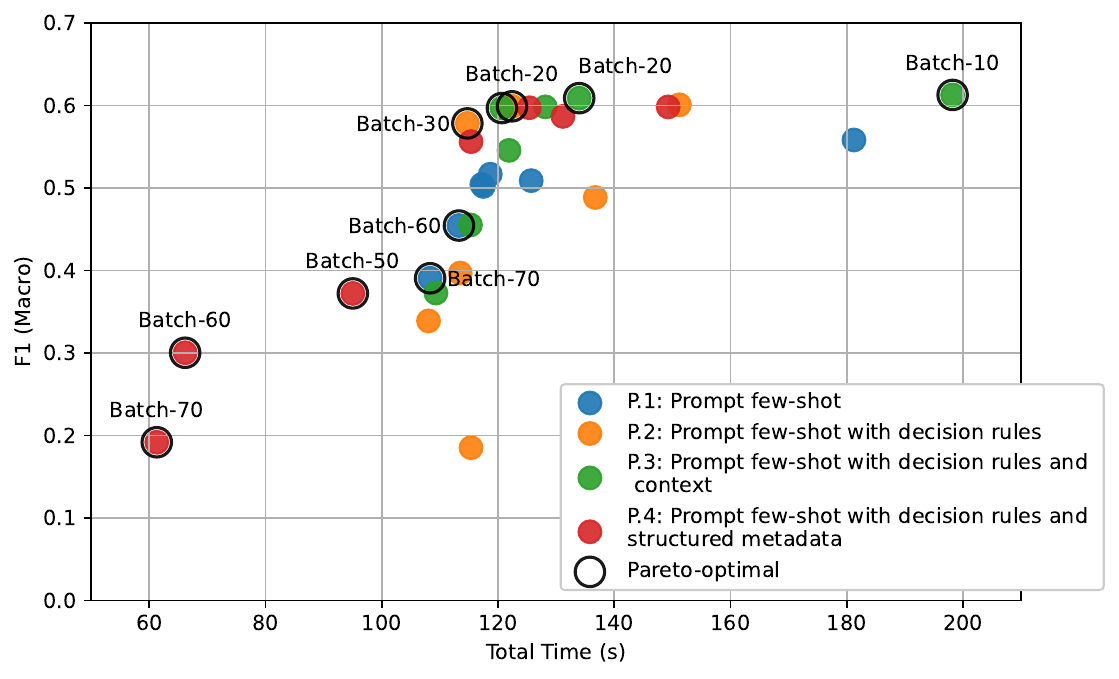} %{figures/pareto_front.pdf}
    \caption{Pareto front. The highlighted points depict the Pareto front for the two dimensions to optimise: Processing Time and Macro-averaged F1 score. }
    \label{fig:pareto_front}
    \Description{This figure shows different configurations and the corresponding Pareto front. The Pareto front represents the dominant configurations under two conflicting objectives, processing time and performance.}
\end{figure*}

The Pareto front is composed of $11$ configurations. Among then, \emph{P4: Prompt few-shot with rules and structured metadata} with batch size $1$, which achieves the highest F1 score. Therefore, if coding performance alone is prioritised, this configuration would be selected. Nevertheless, due to its substantially higher processing time ($1,176.21$ seconds) and energy consumption ($267.85$ Joules), as reported in Sections 4.1 and 4.3, respectively, it is unsuitable for near real-time healthcare debriefing contexts. Excluding this configuration based on feasiblity grounds, we analysed the remaining $10$ configurations that form the Pareto front, as shown in Fig.~\ref{fig:pareto_front} highlighted by black circles: \emph{P.1: Prompt few-shot} with batch sizes $60$ and $70$; two configurations using \emph{P.2: Prompt few-shot with rules} with batch sizes $20$ and $30$; three configurations using \emph{P3: Prompt few-shot with rules and context} with batch sizes $10$, $20$ and $30$; finally, three configurations using \emph{P4: Prompt few-shot with rules and structured metadata} with batch sizes $50$, $60$, $70$.

% Although batch size $1$ configurations are technically non-dominated, they were excluded from practical consideration due to high processing time and energy consumption, as reported in Sections 4.1 and 4.3, respectively, making them unsuitable for near real-time healthcare debriefing contexts. It is noteworthy, if coding performance alone is prioritised, \emph{P4: Prompt few-shot with rules and structured metadata} with batch size $1$ achieves the highest F1 score.

The Pareto analysis revealed distinct deployment preferences. When prioritising computational efficiency, the most suitable configurations were \emph{P4: Prompt few-shot with rules and structured metadata} with batch sizes of $50$, $60$, and $70$. When seeking a balanced trade-off between coding performance and processing time, batch sizes of $20$ and $30$ using rule-based prompts (P.2 and P.3) provide the most promising midpoint.

Given the time-sensitive nature of healthcare simulation debriefings, balanced configurations may be the most appropriate. Among the Pareto-optimal set, the per-label analysis (Table~\ref{tab:per_label_f1_pareto_only}) showed that \emph{P3: Prompt few-shot with rules and context} with batch size $20$ achieved the highest overall label-level performance while maintaining processing time. This configuration, therefore, represents the most favourable trade-off across performance, processing time, and environmental impact.

% \begin{figure*}[t]
%     \centering
%     \begin{subfigure}{0.3\linewidth}
%         \centering
%         \includegraphics[width=1\linewidth]{figures/f1macro_vs_batch.pdf}
%         \caption{F1 (macro) and Batch sizes.}
%         \label{fig:f1macro_vs_batch}
%     \end{subfigure}
%     \hfill
%     \begin{subfigure}{0.3\linewidth}
%         \centering
%         \includegraphics[width=1\linewidth]{figures/time_vs_batch.pdf}
%         \caption{Total Time in seconds and Batch sizes.}
%         \label{fig:placeholder}
%     \end{subfigure}
%     \hfill
%     \begin{subfigure}{0.3\linewidth}
%         \centering
%         \includegraphics[width=1\linewidth]{figures/energy_vs_batch.pdf}
%         \caption{Total System Energy and Batch sizes.}
%         \label{fig:placeholder}
%     \end{subfigure}
%     \caption{Comparisons between performance metrics and batch size for the selected prompts.}
% \end{figure*}

\begin{table}[ht]
\centering
\caption{Per-label F1 scores for Pareto-optimal across prompt versions.}
\label{tab:per_label_f1_pareto_only}
\small
\setlength{\tabcolsep}{4pt}
\begin{tabularx}{\linewidth}{l *{4}{>{\centering\arraybackslash}X}}
\toprule
\multirow{2}{*}{\textbf{Code}} 
& \multicolumn{2}{c}{\textbf{P.2}} 
& \multicolumn{2}{c}{\textbf{P.3}} \\
\cmidrule(lr){2-3} \cmidrule(lr){4-5}
& \textbf{Batch 20} & \textbf{Batch 30} 
& \textbf{Batch 20} & \textbf{Batch 30} \\
\midrule
acknowledging        & 0.72 & 0.70 & \textbf{0.75} & 0.73 \\
escalation           & \textbf{0.70} & 0.55 & 0.65 & 0.64 \\
handover             & 0.15 & \textbf{0.22} & 0.17 & 0.18 \\
questioning          & 0.80 & 0.80 & \textbf{0.82} & 0.80 \\
sharing information  & 0.57 & 0.57 & \textbf{0.58} & \textbf{0.58} \\
task allocation      & 0.66 & 0.62 & \textbf{0.69} & 0.66 \\
\bottomrule
\end{tabularx}
\end{table}

\section{Discussion}

\subsection{Findings by Research Question}
%RQ1: How do prompt design and batching strategies affect processing time when using LLMs for team dialogue coding in healthcare simulation?

With respect to \textbf{RQ1}, which examined whether processing time is affected by prompt design and batching strategies when using LLMs for near real-time dialogue coding in healthcare simulation, our findings showed that both factors substantially influenced processing time, and suggested a batch size range of $n=20$-$70$. In particular, \textbf{prompts incorporating additional context and explicit decision rules, when combined with batching strategies}, led to marked reductions in total processing time. Increasing the number of utterances processed per batch resulted in considerable time savings, a valuable when integrating dialogue analysis into time-sensitive learning activities, where timely feedback is critical for reflective learning, such as post-simulation debriefs \cite{Dieckmann2009,Abulebda2022,Martinez-Maldonado2023tochi}. Our results corroborate previous work on batch prompting \cite{Cheng2023,Ji2025}, confirming that increasing batch size reduces processing time. Extending this line of work, our results show that similar efficiency gains can be achieved when analysing team dialogue in complex, high-stakes contexts such as healthcare simulations.

% However, these processing time gains are accompanied by changes in coding performance, highlighting the need to consider trade-offs across objectives.

%RQ2: How do prompt design and batching strategies affect coding performance when using LLMs for team dialogue coding in healthcare simulation?

Addressing \textbf{RQ2}, we observed that increasing batch size had a noticeable impact on coding performance. \textbf{Batching improved efficiency, but performance (macro-averaged F1) generally decreased as the number of utterances processed simultaneously increased}, mirroring results from Cheng et al., \cite{Cheng2023}. This reveals an inherent trade-off between efficiency and accuracy in near real-time dialogue coding contexts. Among the dialogue codes, \emph{handover} consistently performed worse than the other codes across prompt designs and batch sizes. Results for \emph{handover} detection are consistent with prior BERT-based findings~\cite{Zhao2024BJET}, likely reflecting the deeper conversational context required and conceptual overlap with \emph{sharing information}~\cite{Zhao2024BJET}. Thus, LLMs still struggle to resolve construct-level overlaps—mirroring inconsistencies among human annotators, and echoing educator reports that \emph{handover} may hold limited pedagogical value during debriefing given its overlap with adjacent codes \cite{Zhao2024endusers}. These observations reinforce that coding performance is sensitive to construct interpretation \cite{Liu2025Qualitative}, suggesting that refining coding schemes may improve both model performance and practical relevance—complementing evidence that prompting and batching strategies materially shape LLM coding behaviour \cite{Liu2025Qualitative,SamaraweeraEtAl2026LLMHealthcareSimulation}. Our findings also align with broader claims that LLMs struggle with long-context reasoning, particularly when relevant information is not positioned near the beginning or end of the prompt, suggesting that long, batched prompts may dilute salient signals~\cite{Liu2024lost}. Taken together, these findings suggest that implementing LLMs for team dialogue coding is plausible even in complex, constrained contexts such as healthcare simulation environments.

%RQ3: To what extent do prompt design and batching strate- gies influence the environmental impact of LLM inference for team dialogue coding in healthcare simulation?

Addressing \textbf{RQ3}, we found that energy consumption also decreased as batch size increased, a trend similar to that in RQ1. By empirically demonstrating a near-perfect correlation between processing time and energy consumption, we provide evidence that the \textbf{batching strategy is a direct driver of environmental impact in LLM-based dialogue coding}. This finding establishes that processing time optimisation directly reduces energy demand and associated environmental impact \cite{ding2024sustainable,Elsworth2025,Berthelot2025}. Our results position batching not merely as a performance optimisation technique, but as a practical sustainability lever in LLM coding pipelines \cite{Inie2025CO2STLY}. In doing so, we contribute quantitative evidence that inference energy should not be overlooked and that deployment-level optimisation can meaningfully reduce the environmental foot print \cite{Chien2023reducing}. Furthermore, this study extends the current literature by explicitly quantifying how both prompt design and batching strategies influence environmental impact. We provide empirical justification for using energy consumption as a proxy for environmental impact in local deployment settings, where computational resources are directly tied to institutional infrastructure. To our knowledge, this is the first study to document these relationships within collaborative learning contexts, where local deployment is often necessary due to privacy and infrastructural constraints \cite{Stenseth2025,Algarni2025}.

%RQ4: Which prompt design and batching strategies enable a favourable trade-off between coding performance, processing time, and environmental impact for team dialogue coding in healthcare simulation?

%\cite{Liu2025Qualitative} - use of GPT for for qualitative data analysis - using cloud services - assess three different prompt engineering strategies — Zero-shot, Few-shot, and Few-shot with contextual information — as well as the use of embeddings. GPT-4 had the same difficulty in inferring codes where human coders also struggle. 

Collaborative learning contexts, such as healthcare simulations, are time-constrained and highly sensitive to coding performance, as debriefings often begin immediately after simulation sessions \cite{Martinez-Maldonado2023tochi}. To address this tension, \textbf{RQ4} examined which configurations offered a favourable balance between performance, processing time, and environmental impact. Our results indicate that balanced configurations, specifically those using \textbf{prompts with explicit decision rules, additional conversational context, and moderate batch sizes (approximately $20$ utterances)}, provide the most suitable midpoint. These configurations maintain strong coding performance while substantially reducing processing time and energy consumption, thereby meeting the practical demands of pedagogical instruction, such as debriefing. Our key contribution is to make these trade-offs explicit and empirically quantifiable: prompt design and batching decisions must balance accuracy, efficiency, and sustainability, rather than optimising for performance alone. 

This study extends prior research on LLM-based automated coding, such as Liu et al., \cite{Liu2025Qualitative}, which primarily assessed prompt engineering strategies in terms of coding performance (accuracy). Their findings showed that few-shot prompting improved performance, consistent with our results, which also showed that combining few-shot prompts with explicit decision rules led to accuracy improvements. In contrast, our work incorporates processing time and environmental impact as additional evaluation dimensions and proposes a framework to balance these trade-offs, particularly in local deployment settings critical to  collaborative learning contexts where immediate analysis is required.

Our findings also extend prior feasibility studies of LLM use in \emph{in-situ} simulation settings \cite{SamaraweeraEtAl2026LLMHealthcareSimulation}, which primarily compared prompting strategies based on inference performance. While Samaraweera et al., \cite{SamaraweeraEtAl2026LLMHealthcareSimulation} showed that few-shot prompting improved coding performance and that selectively adding contextual information benefited certain codes accuracy, it also reported substantial increases in environmental impact. We build on this work by evaluating prompting and batching under a unified trade-off framework that integrates performance, processing time, and environmental impact.

% This is more for the Trade-off analysis RQ.
Related work on batch prompting has focused on grouping inference tasks to reduce prompt length. These works were not evaluated in terms of dialogue coding or classification tasks; moreover, they were beyond the scope of healthcare simulation scenarios, nor were they systematically evaluated for multiple conflicting objectives such as accuracy, processing time, and environmental impact.
%In healthcare education contexts, Samaraweera et al. \cite{SamaraweeraEtAl2026LLMHealthcareSimulation} evaluated different prompt designs and coding performance for team coding. Our work goes beyond by systematically evaluating the impact of prompt design and batching strategy on processing time, accuracy, and environmental impact.

% Although overall performance for some codes remained moderate, prior research suggests that even imperfect automated coding can provide valuable support for educators when used as a complementary tool rather than a replacement for expert judgement \cite{Zhao2024endusers,ZambranoEtAl2023FromnCoderToChatGPT,Wang2025BERTvsLLM}. Thus, the configurations identified in this study offer a practical and sustainable pathway toward near real-time dialogue analytics in healthcare simulation.

\subsection{Implications for Learning at Scale}

%This study addresses practical constraints that arise when deploying LLMs in real-world educational settings, particularly in simulation-based healthcare training \cite{Martinez2017,Echeverria2025} and similar face-to-face CSCL situations \cite{Wang2025BERTvsLLM,Martinez2020where,MartinezMaldonado2017DesignEcology} where dialogue coding can potentially be used to support reflection and learning. In such contexts, solutions must operate under strict time and computational limitations while maintaining acceptable levels of accuracy \cite{Martinez-Maldonado2023tochi}. 

This study presents methodological and practical implications for deploying scalable LLM-based dialogue coding in authentic educational settings. Although contextualised in healthcare education, our results apply to other collaborative learning contexts where dialogue is central for learning~\cite{Marshall2024,Papathoma2016}.

From a \emph{methodological standpoint}, this study frames scalable LLM-based dialogue coding in authentic educational settings as a balance between processing time, performance, and environmental impact. Moving beyond prior work that has primarily reported coding accuracy \cite{Liu2025Qualitative,Wang2025BERTvsLLM,Li2025SSRL,Mehta2025}, we show that automated dialogue coding should be engineered and evaluated by considering these three metrics for learning-at-scale analytics. This methodological lens guides configuration choices so they align with the learning activity, deployment constraints, and environmental costs \cite{SamaraweeraEtAl2026LLMHealthcareSimulation,Stenseth2025,Algarni2025,Martinez2017}.
In doing so, it complements research aimed at supporting learning at scale via LLM-driven coding and learner-facing support \cite{Jin2024optimising,Mehta2025,Nie2025,Hutt2024scaling,Moore2024automated} by providing a unified framework for analysing and reporting deployment trade-offs.

%In particular, it offers guidance for face-to-face CSCL contexts \cite{Wang2025BERTvsLLM,Martinez2020where,MartinezMaldonado2017DesignEcology} and simulation-based healthcare training \cite{Martinez2017,Echeverria2025}, where dialogue analysis can support reflection and teamwork learning, but solutions must operate under strict temporal and computational constraints while maintaining reliable coding performance \cite{Martinez-Maldonado2023tochi}. Beyond addressing deployment realities, this work shows how automated dialogue coding can be engineered to balance processing time, performance, and environmental impact for learning-at-scale analytics. 

From a \emph{practical standpoint}, the findings offer guidance for deploying dialogue coding in face-to-face CSCL contexts \cite{Wang2025BERTvsLLM,Martinez2020where,MartinezMaldonado2017DesignEcology} and healthcare simulation-based education \cite{Martinez2017,Echeverria2025}.  Our results indicate that LLM-based analytics can be integrated into time-sensitive educational workflows (e.g., post-simulation debriefings), supporting scalable use of dialogue data to inform reflective discussions and learning. The average processing time was $118.38$ seconds (SD=$25.12$) per session for batch sizes $n > 1$, indicating that the coding can be generated within the timeframe of typical simulation debriefing workflows. Researchers and developers of AI-supported feedback may combine few-shot prompts with additional decision rules, and select batch sizes based on deployment priorities (e.g., coding performance, processing time, and environmental impact). If coding performance alone is priority, a batch size of $1$ is preferable. If the focus is on processing time, larger batch sizes should be selected. For a more balanced trade-off between accuracy and processing time, batch sizes of $n = 20$–$30$ are recommended. Together, they support scalable and context-sensitive integration into collaborative learning practice and contribute to ongoing discussions in CSCL and teamwork science \cite{Salas2018ScienceTeamwork,HmeloSilverJeong2021CSCLMethods}.

\subsection{Limitations and Future Work}

This study evaluated four prompt designs combined with batching strategies that process multiple utterances simultaneously. However, alternative batching approaches were not explored. Future work could investigate distributing classification task across multiple specialised prompts, each identifying a specific dialogue code while processing single utterances. Such task decomposition strategies may yield different trade-offs between performance, efficiency, and energy consumption.
Additionally, the prompts relied on extended few-shot examples and contextual information, resulting in long prompt structures. While effective, these prompts may not be optimally efficient. Future research should explore prompt optimisation techniques, such as structured formats, compressed representations, or schema-based inputs, to reduce token length and improve both computational efficiency and environmental impact.

\subsection{Concluding Remarks}

This study researched how prompt design and batching strategies affect coding performance, processing time, and environmental impact in healthcare simulation dialogue coding. Results show a clear trade-off between efficiency and performance, with rule-based prompts and moderate batch sizes offering the most favourable balance. These findings contribute practical guidance for deploying LLMs in time-sensitive educational settings and highlight the need to optimise configurations for accuracy and sustainability. Future work should explore alternative task decomposition strategies and prompt optimisation techniques to further improve scalability.

\bibliographystyle{splncs04}
\bibliography{bibliography}

\end{document}